\documentclass[journal]{IEEEtran}
\usepackage{todonotes} 

\usepackage{tabularx,adjustbox}
\usepackage{amsmath,amssymb,bm,mathrsfs}
\usepackage{siunitx}

\newcommand{\mi}{\textcolor{black}{j}}

\usepackage{breqn}
\usepackage{amsmath}
\usepackage{graphicx}
\usepackage{mwe}
\usepackage{arydshln}
\usepackage{makecell}
\usepackage{adjustbox}
\usepackage{float}
\usepackage{nomencl}
\makenomenclature
\usepackage{mathtools}
\usepackage{bm}
\usepackage{soul}
\usepackage{siunitx}
\usepackage{cite}
\usepackage{multirow}
\usepackage{arydshln}
\usepackage{xfrac}
\usepackage{nicefrac}
\usepackage{nicematrix,tikz}
\usepackage[T1]{fontenc}
\usepackage{inputenc}
\usepackage{babel}
\usepackage[font=small,labelfont=bf]{caption}
\DeclareSIUnit \var { var } 
\usepackage{titlesec}
\usepackage{etoolbox}
\usepackage{arydshln} 
\usepackage{environ}         
\usepackage{etoolbox}        
\usepackage{graphicx}        
\usepackage{hyperref}
\usepackage{footnote}

\newlength{\myl}
\let\origequation=\equation
\let\origendequation=\endequation

\RenewEnviron{equation}{
  \settowidth{\myl}{$\BODY$}                       
  \origequation
  \ifdimcomp{\the\linewidth}{>}{\the\myl}
  {\ensuremath{\BODY}}                             
  {\resizebox{0.89\linewidth}{!}{\ensuremath{\BODY}}}  
  \origendequation
}
\newcommand{\overbar}[1]{\mkern 1.5mu\overline{\mkern-1.5mu#1\mkern-1.5mu}\mkern 1.5mu}

\def\hlinewd#1{%
\noalign{\ifnum0=`}\fi\hrule \@height #1 %
\futurelet\reserved@a\@xhline}
\makeatother

\begin{document}

\title{General and Unified Model of the Power Flow Problem in Multiterminal AC/DC Networks}
%
%
%

\author{Willem~Lambrichts,~\IEEEmembership{Student Member,~IEEE,}
        and Mario~Paolone,~\IEEEmembership{Fellow,~IEEE,}
\thanks{The authors are with the Swiss Federal Institute of Technology of Lausanne, Switzerland e-mail: (see willem.lambrichts@epfl.ch; mario.paolone@epfl.ch). The project has received funding from the European Union’s Horizon 2020 Research \& Innovation Programme under grant agreement No. 957788.}}%

\maketitle

\begin{abstract}

This paper proposes a generic and unified model of the power flow (PF) problem for multiterminal hybrid AC/DC networks. 
The proposed model is an extension of the standard AC-PF. The DC network is treated as an AC one and, in addition to the \textit{Slack}, \textit{PV} and \textit{PQ} nodes, four new node types are introduced to model the DC buses and the buses connecting the AC/DC interfacing converters (IC). The unified model is solved using the Newton-Raphson method. The extended PF equations can be used in the presence of multiple ICs operating under different control modes. Compared to other recent works, the proposed method allows multiple ICs to regulate the DC voltage simultaneously. This corresponds to more realistic operational conditions that ensure redundancy and allow for more flexible control of the hybrid grid.
The proposed model can be used for networks under unbalanced conditions and allows for an intentionally negative sequence power injection.
In addition to the operational advantages of this method, it is shown that the computational performance of the proposed method is one order of magnitude better than that of other methods presented in the existing recent literature while having the same accuracy.

\end{abstract}



\nomenclature[Aa]{$\mathcal{N}$}{Set of AC nodes}
\nomenclature[Ab]{$\mathcal{M}$}{Set of DC nodes}
\nomenclature[Ac]{$\Gamma$}{Set of IC nodes (i.e. nodes connected to AC or DC side of IC)}

\nomenclature[Ba]{$\overbar{E}$}{Nodal (AC + DC) phase-to-ground voltage phasor\footnote{ \label{note1} For the DC network $\Im \{ \overbar{E}\} = 0 $ and $\Im \{ \overbar{I} \}= 0 $ }}
\nomenclature[Bb]{$\overbar{I}$}{Current phasor (AC + DC)$^{\ref{note1}}$ }
\nomenclature[Bc]{$P$}{Active power}
\nomenclature[Bd]{$Q$}{Reactive power}
\nomenclature[Be]{$\overbar{Y}^{ac}$}{AC compound admittance matrix $ \overbar{I} = \overbar{Y}^{ac} \overbar{E}$}
\nomenclature[Bf]{$Y^{dc}$}{DC compound admittance matrix $ \overbar{I} = Y^{dc}\overbar{E}$}

\nomenclature[Bg]{$I^{sw}$}{DC current modelling IC's switching losses}
\nomenclature[Bh]{$\overbar{E}^{c}$}{AC voltage drop modelling IC's conduction losses}
\nomenclature[Bi]{$S^{loss}$}{Total power losses over IC}

\nomenclature[Bj]{$R_{eq}$}{Equivalent resistance of the IGBT}
\nomenclature[Bk]{$T_{ON}$}{Equivalent time constants of transistor's turn-on}
\nomenclature[Bl]{$T_{OFF}$}{Equivalent time constants of transistor's turn-off}
\nomenclature[Bm]{$T_{REC}$}{Equivalent time constants of diode's reverse recovery}
\nomenclature[Bn]{$\overbar{Z}^{filter}$}{Impedance of the filter of IC}

\nomenclature[Bo]{$\mathbf{J}$}{Jacobian matrix}
\nomenclature[Bp]{$\mathbf{x}$}{Vector of the unknown variables}
\nomenclature[Bq]{$\mathbf{y}$}{Vector of the mismatches}
\nomenclature[Br]{$\epsilon$}{Convergence criteria}

\nomenclature[Pa]{$\bullet_i$}{Subscript indicating the AC nodes}
\nomenclature[Pb]{$\bullet_j$}{Subscript indicating the DC nodes}
\nomenclature[Pc]{$\bullet_l$}{Subscript indicating node at AC side of IC}
\nomenclature[Pd]{$\bullet_k$}{Subscript indicating node at DC side of IC}

\nomenclature[Pe]{$\bullet^{\prime}$}{Real part of complex number}
\nomenclature[Pf]{$\bullet^{\prime \prime}$}{Imaginary part of complex number}
\nomenclature[Pg]{$\bullet^{\phi}$}{Superscript for phase angle $\phi \in \{a,b,c \}$}
\nomenclature[Ph]{$\bullet^{\ast}$}{Superscript for reference point or setpoint}
\nomenclature[Pi]{$\bullet^{0}$}{Superscript indicating zero sequence}
\nomenclature[Pj]{$\bullet^{+}$}{Superscript indicating positive sequence}
\nomenclature[Pk]{$\bullet^{-}$}{Superscript indicating negative sequence}

\printnomenclature

\section{Introduction}

Multiterminal hybrid AC/DC networks are gaining more interest nowadays in the area of HVDC as well as in hybrid AC/DC microgrids. Indeed, multiterminal HVDC systems allow the interconnection of transition networks to increase the flexibility and integration of renewables \cite{ENTSOE2019}. On the other hand, hybrid AC/DC microgrids are a promising solution to increase the share of distributed generation in future power grids that are expected to massively rely on converter-interfaced renewable resource generation \cite{eghtedarpour2014power}. 

Power flow (PF) studies are a crucial element in the analysis, planning and operation of these modern power systems that are transitioning to hybrid AC/DC systems.
From a modelling point of view, the PF analysis of multiterminal HVDC and microgrids is identical. The main challenge comes from the incorporation of the AC/DC Interfacing Converters (IC). Indeed, the IC can operate using different control modes that affect the PF model. Typically, in a Voltage Source Converter (VSC), the d- and q-components of the current and voltages are decoupled. This allows for the control of two variables simultaneously, i.e. active power, reactive power, AC voltage, or DC voltage. The common controllable pairs of variables are: $P_{ac} - Q_{ac}$, $P_{ac} - \lvert E_{ac} \rvert $, $E_{dc} - Q_{ac}$ or $E_{dc} - \lvert E_{ac} \rvert $. For the first two operating modes, $P_{ac} - Q_{ac}$ and $P_{ac} - \lvert E_{ac} \rvert $, the connected AC bus is considered as a \textit{PQ node} and \textit{PV node}\footnote{We use the standard PF nomenclature where in a \textit{PV node} the active power and the voltage magnitude are controlled and in a \textit{PQ node} the active and reactive power are controlled}. Therefore, the traditional AC power flow theory can be applied. Furthermore, using the active power balanced and IC loss model, the DC side is modelled as a \textit{P-node} and can be easily included in the PF model. The main challenge in the construction of a unified PF comes from the operating modes where the IC controls the DC voltage (that is, $E_{dc} - Q_{ac}$ and $E_{dc} - \lvert E_{ac} \rvert $ ). Because the DC voltage is a control variable, only one AC variable is controlled (either $Q_{ac}$ or $\lvert E _{ac} \rvert$), and the traditional PF theory cannot be used anymore. Note that at least one IC is required to control the DC voltage to ensure the stability of the DC grid. 

Furthermore, hybrid AC/DC microgrids are often subjected to strong unbalanced loading conditions. Therefore, it is in the interest of the system's quality of supply to have a PF model that allows to consider generic AC unbalances including, but not limited to, the intentional injection of negative sequence power.

The structure of the paper is as follows: Section \ref{sec:stateoftheart} gives a review of the state-of-the-art on PF methods for hybrid AC/DC networks and discusses several major limitations of these methods that our proposed model tackles. Section \ref{Sec:Methodology} presents the proposed generic hybrid AC/DC PF model. The load flow equations are presented for all different node types and operation modes of the ICs. Furthermore, a detailed loss model is presented that improves the accuracy of the PF solution. In Section \ref{Sec:Casestudy}, a case study is presented to numerically validate the proposed method on a hybrid AC/DC microgrid under balanced and unbalanced loading conditions. Section \ref{sec:comparison} presents an in-depth comparison and benchmark with a publicly available MATPOWER-based PF algorithm for hybrid AC/DC networks \cite{zimmerman2010matpower}. The computation time is analysed for multiple hybrid AC/DC networks, including a large network to show the scalability of our proposed method.

\section{State-of-the-art review} 
\label{sec:stateoftheart}

The PF problem for AC/DC networks has been studied extensively since the 1980s. Reference \cite{braunagel1976inclusion} proposed a unified PF model that includes the DC network model and allows for a multiterminal configuration. The model is solved using the Newton-Raphson (NR) method. \cite{smed1991new} presented a method for a decoupled PF where the ICs are modelled as voltage-dependent loads to eliminate the need for DC variables. However, these methods are only valid for Line-Commutated Converters (LCC). For VSCs, the mathematical model is fundamentally different, and the above-mentioned methods cannot be used anymore.

In the applications of VSCs, the PF models proposed in the literature can be classified into two main groups: \textit{sequential} and \textit{unified models}. \\

In the \textit{sequential models}, the DC grid and AC grid are solved separately and iteratively linked using the active power balance at the IC controlling the DC voltage \cite{beerten2010sequential}. Notice that all, except for one, ICs are assumed to operate in PV or PQ mode. Therefore, the traditional PF theory is used to model the AC grid. The active power injection of the IC controlling the DC voltage is unknown and is computed using the active power balance between the AC and DC network.
Reference \cite{liu2010improved} proposes an improved sequential PF method that includes the IC's tap positions as an additional state variable to enhance the robustness of the PF calculation.

The main problem with these methods is the need for an iterative procedure with multiple computation loops, which increases the computational complexity significantly. \\

In the \textit{unified models}, the AC and DC power flows are solved as one problem by modelling the entire network (i.e., AC network, DC network and IC) as one system. 
In the specific case where the DC voltage is regulated using droop control, references \cite{aprilia2017unified, nguyen2019power, chai2016unified, eajal2016unified} propose a method where the $V_{dc} - P$ droop curves are incorporated directly into the PF model. 

When DC voltage is regulated in an optimal manner, not through droop control, only a few unified approaches have been suggested. 
Reference \cite{baradar2011modeling} proposes a unified model based on the sequential approach of \cite{beerten2010sequential}. The updated unified method includes the converter losses of the DC voltage-controlling IC as an optimisation variable. Using this additional variable, the active power injection on the AC side of this IC is equal to the sum of all DC power injections and losses. 

The authors of \cite{acha2013new} propose a new equivalent representation of a VSC where the converter is modelled as an ideal tap-changing transformer with a complex tap. The tap magnitude corresponds to the VSC's modulation index and the tap angle is equal to the phase angle of the AC voltage at the IC's node. An additional shunt susceptance and resistance are included to model the reactive power flow and the converter's losses. The approach allows to describe the IC's fundamental frequency operation as a two-port model. \cite{acha2016generalized} includes the two-port model in a unified PF model that can handle the different operating modes and is solved using the NR method. The method requires new additional control variables and can only model the positive sequence operation of ICs.

The authors in \cite{alvarez2021flexible, alvarez2021universal} propose a PF method using the Flexible Universal Branch Model (FUBM). The FUBM is based on the above-described two-port model and can realistically model AC transmission lines and VSCs operating under different operation modes. The FUBM PF-based method has been made publicly available as an extension of the MATPOWER tool and is used to benchmark our proposed method in Section \ref{sec:comparison}. References \cite{rezvani2021unified, rezvani2021generalized} follow a similar approach in which the AC/DC IC is represented as a two-port model and is included as a building block of the compound admittance matrix of the entire AC/DC network. Therefore, the method can be integrated into conventional PF programmes with minimal modifications. Reference \cite{kumar2021newton} proposes an AC-equivalent approach in which every DC line is replaced by a set of parallel AC lines and in which the ICs are replaced by an equivalent line model dependent on the modulation index.

Reference \cite{renedo2019simplified} proposes another method in which every IC is modelled using two conventional AC generators, one for the AC side and one for the DC side, and coupled by a linear constraint to ensure energy conservation. Furthermore, the DC network is modelled as an AC one, so existing AC-PF tools can be reused. \\

The main limitation of all methods proposed previously in the literature is that only one IC can regulate the DC voltage. 
When multiple ICs regulate the voltage of the DC grid, the problem becomes unfeasible and does not converge to a solution. This fundamental limitation has been identified in \cite{baradar2011modeling, alvarez2021flexible}. 
Having multiple ICs controlling $V_{dc}$ is crucial for numerous reasons. 
1) The security of supply of the DC system since a redundant number of converters can better keep the DC voltage within nominal bounds. Therefore, when one converter goes offline, the other converters will continue to maintain the DC voltage level.
2) When multiple converters control the DC voltage level, the power required to maintain the nominal DC voltage setpoint is shared over multiple converters. Therefore, more power can be exchanged between AC and DC networks, allowing for a broader range of operations than when only a single IC can control the DC voltage. 
3) When DC transformers are present, the DC grid can have multiple voltage levels, which leads to a more optimal control of the entire grid \cite{barcelos2022direct}. The hybrid AC/DC grid used for the validation of our proposed LF method is a real grid available at the EPFL campus that has multiple DC transformers and is inspired by real microgrid benchmarks. The methods previously presented in the literature are not usable in such a network.
An additional major limitation of the methods proposed in the literature is that microgrids are often subjected to unbalanced loading conditions. These are created by, e.g. single-phase photovoltaic inverters or electric vehicle chargers. None of the models can handle these unbalanced conditions. 
Furthermore, the proposed PF models cannot account for the intentional injection of negative sequence power that is often required to compensate for the unbalanced loading conditions. 
The final element of the proposed unified PF model is to improve the computational speed of the tool in comparison to existing ones. \\

In this respect, this paper proposes a generic method that tackles these four fundamental limitations at the same time. The method is based on the AC-PF and is suitably extended to include the DC network and the ICs. The DC network is treated as a standard AC one and the ICs are treated in a generic way. Depending on their operation mode, i.e. if a voltage or power reference is tracked, the PF equations are suitably adapted. The DC voltage control is no longer limited to only one IC. 
The proposed method can be used for all types of hybrid networks (i.e., multiterminal HVDC or hybrid AC/DC microgrids) under balanced or unbalanced conditions. Furthermore, the method allows us to accurately model the AC/DC grid when a negative sequence power is injected to compensate for unbalances. \footnote{The source code is made publicly available on \url{https://github.com/DESL-EPFL}}.

\section{Methodology}
\label{Sec:Methodology}
\subsection{Node types in hybrid AC/DC networks}

The PF problem requires an exact model of the AC network, the DC network, and the ICs.
The AC network consists of three types of buses: \textit{Slack}, \textit{PV} and \textit{PQ} nodes and is modelled using the standard PF theory. 
The DC network is modelled identically to the AC network with $Q = 0$ and $\overbar{Z} = R$ in order to reuse the AC-PF theory. 
Because of the nature of the ICs, which are typically VSCs, it is not possible anymore to use the traditional PF theory and an extension is needed where the model equations are dependent on the converter's operational mode. Typically, the control architecture of VSCs allows the control of two variables simultaneously: $E_{dc}-Q_{ac}$, $P_{ac}-Q_{ac}$ and $ E_{dc}- \lvert E_{ac} \rvert$. 
Tab. \ref{NodeTypes} gives an overview of all possible node types in hybrid AC/DC grids. Note that at least one VSC is required to impose the DC voltage ($E_{dc}$) \cite{baradar2011modeling}. 
\renewcommand{\arraystretch}{1.2}

\normalsize
\renewcommand{\arraystretch}{1.5}
\begin{table}[!h]
\caption{Different types of nodes in hybrid AC/DC networks and their known and unknown variables.}
\resizebox{1.05\columnwidth}{!}{%
\begin{tabular}{lllll}
\textbf{Bus Type   }             & \textbf{VSC contrl. }                     & \textbf{Known var. }                                             & \textbf{Unknown var. } & \textbf{Index}\\ \hline
AC slack                   &                                  & $\lvert E_{ac} \rvert$, $\angle E_{ac}$                               & $P_{ac}$,$Q_{ac}$       & $s \in \mathcal{N}_{slack}$   \\ \hline
$P_{ac}$, $Q_{ac}$                 &                                  & $P_{ac}$,$Q_{ac}$                                                & $\lvert E_{ac} \rvert$, $\angle E_{ac}$    & $i \in \mathcal{N}_{PQ}$            \\ \hline
$P_{ac}$, $\lvert E_{ac} \rvert$                 &                                  & $P_{ac}$,$\lvert E_{ac} \rvert$                                                & $Q_{ac} $, $\angle E_{ac}$  & $i \in \mathcal{N}_{PV}$              \\ \hline
\multirow{2}{*}{$VSC_{ac}$} & $P_{ac}$ - $Q_{ac} $                & $P_{ac}$ $Q_{ac}$     & $\lvert E_{ac} \rvert$, $\angle E_{ac}$           &  $l \in \Gamma_{PQ}$  \\ \cdashline{2-5}
                        & $E_{dc}$ - $Q_{ac}  $                   & $Q_{ac}$              & $P_{ac}$ $\lvert E_{ac} \rvert$ $\angle E_{ac}$     &  $l \in \Gamma_{E_{dc}Q}$        \\ \cdashline{2-5} 
                        & $P_{ac}$ - $\lvert E_{ac} \rvert$ & $P_{ac}$ $\lvert E_{ac} \rvert$                           & $Q_{ac}$ $\angle E_{ac}$  &   $k \in \Gamma_{PV}$            \\ \hline
\multirow{2}{*}{$VSC_{dc}$} & $P_{ac}$ - $Q_{ac}$                 & $P_{dc}$              & $E_{dc}$           &  $k \in \Gamma_{PQ}$   \\ \cdashline{2-5}
                        & $E_{dc}$ - $Q_{ac} $                    & $E_{dc}$              & $P_{dc}$           &   $k \in \Gamma_{E_{dc}Q}$  \\  \cdashline{2-5} 
                        & $P_{ac}$ - $\lvert E_{ac} \rvert$ & $P_{dc}$      & $E_{dc}$  &   $k \in \Gamma_{P V}$            \\ \hline
$P_{dc}$                     &                                  & $P_{dc}$                                                    & $E_{dc}$          & $j \in \mathcal{M}_{P}$    \\ \hline
$E_{dc}$                     &                                  & $E_{dc}$                                                    & $P_{dc}$           & $j \in \mathcal{M}_{V}$   \\ \hline
\end{tabular} \label{NodeTypes}
}
\end{table}
\normalsize

\subsection{Power flow equations}
\begin{figure}[!h]
\centering
  \includegraphics[width=0.99\linewidth]{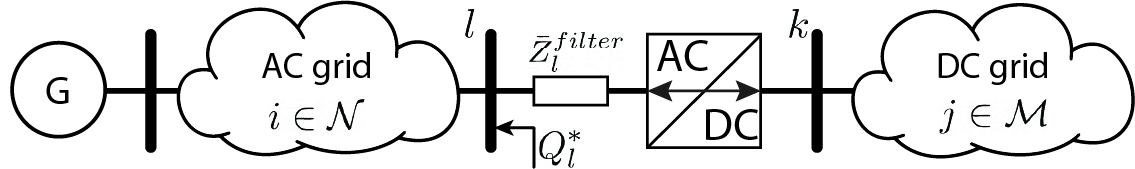}
  \caption{The generic hybrid AC/DC network. Only one AC/DC converter is displayed for simplicity.}
  \label{gengrid}
\end{figure}

A generic hybrid AC/DC network is considered with $\mathcal{N}$ AC nodes and $\mathcal{M}$ DC nodes, where buses $(l, k) \in \Gamma$ are the couples of AC/DC converter buses (see Fig.\ref{gengrid}). Furthermore, we assume $l \in \mathcal{N} $ and $k \in \mathcal{M}$. 
Therefore, $(\mathcal{N} = \mathcal{N}_{slack} \cup \mathcal{N}_{PQ} \cup \mathcal{N}_{PV} \cup \Gamma_l )$ and $(\mathcal{M} = \mathcal{M}_{P} \cup \mathcal{M}_{V} \cup \Gamma_k )$

The two grids are interlinked by one or more interfacing converters (i.e., $\lvert \Gamma \rvert \geq 1$) that can operate under different control modes. Notice that the PF equations below are written in rectangular coordinates\footnote{This was a design choice by the authors, however, the model can also be easily described in polar coordinates.}.

\subsubsection{AC network}

The PF equations for \textit{PQ} nodes read:
\begin{eqnarray}
        &\Re \Big\{ \overbar{E}_{i}^{\phi} \sum\nolimits_{n \in \mathcal{N}} \underbar Y_{i,n}^{ac} \underbar E_{n}^{\phi} \Big\} = P^{\phi \ast}_{i} , \quad \text{for } i \in \mathcal{N}_{PQ} \label{PQ_p}\\ 
        &\Im \Big\{ \overbar{E}_{i}^{\phi} \sum\nolimits_{n \in \mathcal{N}} \underbar Y_{i,n}^{ac} \underbar E_{n}^{\phi} \Big\} = Q^{\phi \ast}_{i} , \quad \text{for } i \in \mathcal{N}_{PQ} \label{PQ_q}
\end{eqnarray}

The PF equations for \textit{PV} nodes read:
\begin{eqnarray}
        \Re \Big\{ \overbar{E}_{i}^{\phi} \sum\nolimits_{n \in \mathcal{N}} \underbar Y_{i,n}^{ac} \underbar E_{n}^{\phi} \Big\} &&= P^{\phi \ast}_{i} , \quad \text{for } i \in \mathcal{N}_{PV} \label{PV_p}\\ 
         {E_{i}^{\phi \prime}}^2 + {E_{i}^{\phi \prime \prime}}^2 &&= {E_{i}^{\phi \ast}}^2 , \quad \text{for } i \in \mathcal{N}_{PV} \label{PV_v}
\end{eqnarray}
where $P^{\phi \ast}_i$ and $Q^{\phi \ast}_i$ are the active and reactive power nodal injections at node $i$ and phase $\phi \in \{a,b,c \}$. 
The $\underbar{$\bullet $}$ indicates the complex conjugate, the apostrophes $ \bullet^{\prime} $ and $ \bullet^{\prime \prime} $ 
refer to the real and imaginary parts of the phase-to-ground voltage $\overbar{E}_{i}$. $\overbar{Y}^{ac}$ is the admittance matrix of the AC network.

\subsubsection{DC network}
The PF equations for \textit{P} nodes in the DC grid read: 
\begin{eqnarray}
    & E_{j} \sum\nolimits_{m \in \mathcal{M}} Y_{j,m}^{dc} E_{m}  = P^{\ast}_{j} , \quad \text{for } j \in \mathcal{M}_{P} \label{P_dc}
\end{eqnarray}

The PF equations for \textit{V} nodes in the DC grid read: 
\begin{eqnarray}
    & E_{j} = E^{\ast}_{j} , \quad \text{for } j \in \mathcal{M}_{V} \label{V_dc}
\end{eqnarray}

\subsubsection{VSC interfacing converters}

For the $\mathbf{E_{dc} - Q_{ac}}$ operating mode, the extended PF equations are based on the power balance \eqref{Pblanced}. Notice that the reactive power losses over the filter are already accounted for in the control of the VSC (see Fig.\ref{gengrid}).
\begin{flalign}
    &P_{l}^a + P_{l}^b + P_{l}^c + P^{loss}_{(l,k)} + P^{filter}_{(l,k)}  = P_{k},   \nonumber \\ 
    &Q_{l}^a + Q_{l}^b + Q_{l}^c - Q^{loss}_{(l,k)}   = Q_{l}^{\ast}, \quad \quad \quad \text{for } (l,k) \in \Gamma_{E_{dc} Q}
    \label{Pblanced}
\end{flalign}
with $Q_{l}^{\ast}$ the reactive power setpoint of the VSC and subscripts $\bullet _k$ and $\bullet _l$ referring to the resp. DC and AC side of the VSC.
In a balanced system, the power is shared equally among the three phases, and thus $P_{l}^{\phi} = \frac{1}{3} P_{k}$ and $Q_{l}^{\phi} = \frac{1}{3} Q_{l}^{\ast}$.  
In unbalanced systems, the phase-locked loop (PLL) usually synchronizes to the AC network positive sequence, therefore, only the positive sequence powers are injected:
\begin{flalign}
&\begin{cases}
      P_{l}^0  & \ = 0\\
      P_{l}^+  &  +  P^{+ loss}_{(l,k)} + P^{+ filter}_{(l,k)} = P_{k}\\
      P_{l}^-  & \ = 0
    \end{cases}   \nonumber \\ 
    \nonumber\\
&\begin{cases}
      Q_{l}^0 & \ = 0\\
      Q_{l}^+ & - Q^{+ loss}_{(l,k)}  = Q^{\ast}_{l}\\
      Q_{l}^- & \ = 0
    \label{Punblanced}
\end{cases}  &&
\end{flalign}
where
\begin{eqnarray}
P_{l}^{+} & = & \Re \Big\{ \overbar{E}_{l}^+ \sum\nolimits_{n \in \mathcal{N}} \underbar Y_{(l,n)}^{ac} \underbar E_{n}^+ \Big\}, \label{Pl}\\
Q_{l}^{+} & = & \Im \Big\{ \overbar{E}_{l}^+ \sum\nolimits_{n \in \mathcal{N}} \underbar Y_{(l,n)}^{ac} \underbar E_{n}^+ \Big\}, \label{Ql}\\
P_{k}^{+} & = & \Re \Big\{ {E}_{k} \sum\nolimits_{m \in \mathcal{M}} Y_{(k,m)}^{dc}  E_{m} \Big\}  \label{Pk}
\end{eqnarray}
and $E_k = E^{\ast}$ is the DC voltage setpoint.

\vspace{0.5em}
Because $\overbar{S}_{l}^0 = P_{l}^0  + \mi Q_{l}^0 = \overbar{E}^0_{l} \underbar{I}^0_{l} $, the homopolar component of the voltage ${E}^0_{l}$, or the current $I^0_{l}$, has to be zero. For a VSC, where the voltage is controlled, ${E}^0_{l}$ is set to zero. For a current source converter, $I^0_{l}$ would be zero (idem for the negative sequence component). 
Note here that this operation distinction has to be made, since the expression for the homopolar and negative sequence powers in \eqref{Punblanced}  cannot be used. Using these power injections results in a trivial expression, and thus in an underdetermined problem. 
\vspace{0.5em}

Substituting expressions \eqref{Pl}, \eqref{Ql} and \eqref{Pk} into \eqref{Punblanced} reads:
\begin{flalign}
&\begin{cases}
      E^{0  \prime}  = 0 \\
      \Re \Big\{ \overbar{E}_{l}^+ \sum\nolimits_{n \in \mathcal{N}} \underbar Y_{(l,n)}^{ac} \underbar E_{n}^+ \Big\}  +  P^{+ loss}_{(l,k)} + P^{+ filter}_{(l,k)} = \\ 
      \qquad \qquad \quad E_{k}^{\ast} \left( Y_{(k,k)}^{dc} E_{k}^{\ast} + \sum\nolimits_{\substack{m \in \mathcal{M}\\
                  m \neq k}} Y_{(k,m)}^{dc} E_{m} \right) \\
      E^{-  \prime }  = 0    
\end{cases}    \label{VSC_p} \\
 \nonumber \\
&\begin{cases}
      E^{0  \prime  \prime}  = 0\\
      \Im \Big\{ \overbar{E}_{l}^+ \sum\nolimits_{n \in \mathcal{N}} \underbar Y_{(l,n)}^{ac} \underbar E_{n}^+ \Big\} - Q^{+ loss}_{(l,k)} = Q^{\ast}_l\\
      E^{-  \prime  \prime}  = 0 \label{VSC_q}
\end{cases}  
\end{flalign}

Rewriting the positive sequence component of \eqref{VSC_p} to $E_k^{\ast}$ (the controllable DC voltage) leads to the quadratic equation \eqref{Edc_quadratic}:

\begin{eqnarray} \label{Edc_quadratic}
\Big(  Y_{(k,k)}^{dc}  \Big) {E_{k}^{\ast}} ^2 + \Big( \sum\nolimits_{\substack{m \in \mathcal{M} \\
                  m \neq k}} Y_{(k,m)}^{dc} E_{m} \Big) E_{k}^{\ast} \nonumber \\
 \qquad  \qquad   \qquad  - \Re \Big\{ \overbar{E}_{l}^+ \sum\nolimits_{n \in \mathcal{N}} \underbar Y_{(l,n)}^{ac} \underbar E_{n}^+ \Big\} = 0
\end{eqnarray}

Solving the quadratic equation \eqref{Edc_quadratic} to $E_{k}^{\ast}$ results in two solutions: one close to 1 p.u. and another (infeasible) close to 0 p.u. Because the operational voltage of a grid is close to 1 p.u., the only feasible solution is given by the positive equation \eqref{Edc_final}.

\small
\begin{flalign} \label{Edc_final}
E^{\ast}_{k} = &-\frac{\sum\nolimits_{\substack{m \in \mathcal{M}\\
                  m \neq k}} Y_{(k,m)}^{dc} E_m}{2 Y_{(k,k)}^{dc}}  \pm && \\ \nonumber
&  \frac{\sqrt{\left(\sum\nolimits_{\substack{m \in \mathcal{M}\\
                  m \neq k}} Y_{(k,m)}^{dc} E_m \right)^2 - 4 Y_{(k,k)}^{dc} \Re \Big\{ \overbar{E}_{l}^+ \sum\nolimits_{n \in \mathcal{N}} \underbar Y_{(l,n)}^{ac} \underbar E_{n}^+ \Big\} }}{2 Y_{(k,k)}^{dc}}&&
\end{flalign}
\normalsize

Notice that \eqref{Edc_final} is dependent on the positive sequence nodal voltage component $\overbar{E}^+$. Using the standard Fortescue symmetrical component decomposition, the nodal voltage can be transformed back to the phase domain:
    \begin{equation}
            \label{symmetrical}
            \left[\begin{IEEEeqnarraybox*}[][c]{,c,}
            \overbar{E}^{0 }  \\
            \overbar{E}^{+ }  \\
            \overbar{E}^{- } 
            \end{IEEEeqnarraybox*}\right]
          = \frac{1}{3} \left[\begin{IEEEeqnarraybox*}[][c]{,c/c/c,}
            1 & 1 & 1 \\
            1 & \overbar{\alpha} & \ \overbar{\alpha}^2\\
            1 & \ \overbar{\alpha}^2 & \overbar{\alpha}
            \end{IEEEeqnarraybox*}\right]
        \cdot 
            \left[\begin{IEEEeqnarraybox*}[][c]{,c,}
            \overbar{E}^{a} \\
            \overbar{E}^{b} \\
            \overbar{E}^{c} 
            \end{IEEEeqnarraybox*}\right]
    \end{equation}

For the $\mathbf{P_{ac} - Q_{ac}}$ operating mode, the AC side of a generic VSC, which can inject positive and negative power references, is modelled as \eqref{VSC2_p} and \eqref{VSC2_q}. In the case that the IC only injects the positive sequence, the power balance of the negative sequence is replaced by $\overbar{E}^{-}  = 0$.

\begin{flalign}
 &\begin{cases}
      E^{0  \prime }  = 0\\
      \Re \Big\{ \overbar{E}_{l}^+ \sum\nolimits_{n \in \mathcal{N}} \underbar Y_{(l,n)}^{ac} \underbar E_{n}^+ \Big\}  - P_{(l,k)}^{+ loss} = P^{+ \ast}_l\\
       \Re \Big\{ \overbar{E}_{l}^- \sum\nolimits_{n \in \mathcal{N}} \underbar Y_{(l,n)}^{ac} \underbar E_{n}^- \Big\}  - P_{(l,k)}^{- loss} = P^{- \ast}_l 
    \end{cases}       \label{VSC2_p} \\
     &\begin{cases}
      E^{0  \prime  \prime}  = 0\\
      \Im \Big\{ \overbar{E}_{l}^+ \sum\nolimits_{n \in \mathcal{N}} \underbar Y_{(l,n)}^{ac} \underbar E_{n}^+ \Big\} - Q^{+ loss}_{(l,k)} = Q^{+ \ast}_l\\
      \Im \Big\{ \overbar{E}_{l}^- \sum\nolimits_{n \in \mathcal{N}} \underbar Y_{(l,n)}^{ac} \underbar E_{n}^- \Big\} - Q^{- loss}_{(l,k)} = Q^{- \ast}_l,  \label{VSC2_q}
    \end{cases} \\
    & \qquad \qquad \qquad  \qquad \qquad \qquad    \qquad \qquad \text{for } (l,k) \in \Gamma_{PQ} \nonumber
\end{flalign}

The active power balance also imposes the DC power injection on the DC side. Taking the converter and filter losses into account gives:
\begin{eqnarray}
    P^{\ast}_l && = E_{k} \sum\nolimits_{m \in \mathcal{M}} Y_{k,m}^{dc} E_{m} - P^{loss}_{(l,k)} - P^{filter}_{(l,k)} 
    \label{VSC2_pdc}
\end{eqnarray}

For the $\mathbf{E_{dc} - \lvert E_{ac} \rvert}$ operating mode, the interfacing VSC model consists of equations \eqref{VSC_p} and \eqref{PV_v}.

\subsection{AC/DC interfacing converter loss model}

\begin{figure}[!h]
\centering
  \includegraphics[width=0.8\linewidth]{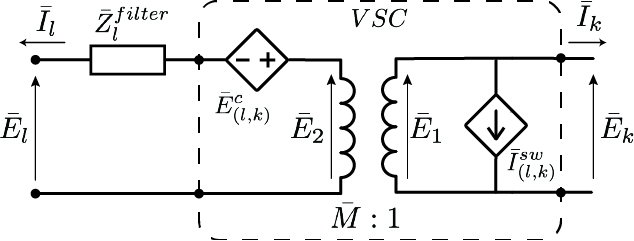}
  \caption{Transformer-like model of an inverter leg of the AC/DC interfacing converter with RL filter.}
  \label{fig:AFE_loss}
\end{figure}

The losses of the AC/DC interfacing converter can be included for a more accurate grid model. Assuming that the semiconductor switches of the VSCs are Insulated-Gate Bipolar Transistors (IGBT), references \cite{scapino2003transformerlike, scapino2005vsi} propose an accurate VSC loss model.  The converter model that includes the losses and filter is shown in Fig.\ref{fig:AFE_loss}.
The conduction losses are modelled as a voltage source $\overbar{E}_{l}^c$ on the AC side \eqref{VCphasorReq} and the switching losses as a DC current source $I_{k}^{sw}$ \eqref{ISphasor}. 

 \begin{eqnarray}
        \overbar{E}^{c}_{(l,k)} =  R_{eq} \left(\lvert \overbar{I}_{l} \rvert \right) \overbar{I}_{l} \label{VCphasorReq}
\end{eqnarray}
\begin{eqnarray}
    \overbar{I}^{sw}_{(l,k)} = 2 \text{\footnotesize  $  \frac{T_{ON} + T_{OFF} + T_{REC}}{T_s} $ } \frac{1}{N}  \text{cot}(\frac{\pi}{N})  \lvert \overbar{I}_{l} \rvert  \label{ISphasor}
 \end{eqnarray}

where $R_{eq}$ is the equivalent resistance of the IGBT, $T_s$ is the switching period, and $N = \nicefrac{f_{s}}{f_{line}} $ is the ratio between the switching frequency and the line frequency. $T_{ON}$ and $T_{OFF}$  are the equivalent time commutation constants characterising the transistor's turn-on and turn-off effects under the test conditions, and $T_{REC}$ is the reverse recovery at turn-off of the diode. It can be observed that the losses, $\overbar{E}_{l}^c$ and $I_{k}^{sw}$, are only dependent on the converter's switching frequency and the several parameters that can be easily derived from the semiconductor's datasheet.

The total losses are computed as the sum of the conduction and switching losses:
    \begin{eqnarray}
    \overbar{S}^{loss}_{(l,k)}  = \overbar{E}_{(l,k)}^c \sum\nolimits_{n \in \mathcal{N}} \underbar{Y}^{ac}_{(l,n)} \underbar{E}_n  + I_{(l,k)}^{sw} E_{k} \label{Plossfirst}
    \end{eqnarray}

The active losses over the \textit{RL}-filter with impedance $\overbar{Z}^{filter}_l$ can be straightforwardly modelled as:
\begin{equation}
    P^{filter}_{(l,k)} = \Re \Big\{ \overbar{Z}^{filter}_l \Big\lvert \sum\nolimits_{n \in \mathcal{N}} \overbar{Y}^{ac}_{(l,n)} \overbar{E}_n \Big\rvert^2 \Big\}
\end{equation}

\subsection{Solution via the Newton-Raphson method}

The PF problem given by \eqref{PQ_p}-\eqref{PV_v}, \eqref{P_dc}, \eqref{V_dc}  \eqref{VSC_p}, \eqref{VSC_q}, the positive solution of \eqref{Edc_final}, \eqref{VSC2_p} and \eqref{VSC2_q}
is solved in a unified way via the Newton-Raphson (NR) method. In compact matrix formulation, it can be written as:
\begin{eqnarray}
    &&\mathbf{J}^{(\nu)} \cdot  \Delta \mathbf{x}^{(\nu + 1)} = \Delta \mathbf{y}^{(\nu)} \\
    &&\text{where} \nonumber \\ 
    && \qquad \Delta \mathbf{x}^{(\nu +1)} = \mathbf{x}^{(\nu +1)} -\mathbf{x}^{(\nu)}\\
    && \qquad \Delta \mathbf{y}^{(\nu)} = \mathbf{y}^{\ast} - F(\mathbf{x}^{(\nu)})
\end{eqnarray}
where, $\nu$ is the iteration's step, $\mathbf{J}$ is the PF Jacobian composed by the first-order partial derivatives of the PF model, $\mathbf{x}$ is the vector of the unknown variables. $\Delta \mathbf{y}$ is the mismatch related to the known variables, i.e., the difference between the power or voltage setpoint, indicated by the symbol $\ast$, and the evaluated PF equations. 
For the hybrid system discussed in the case study, the linearised system of equations gives:
    \renewcommand{\arraystretch}{1.15}
    \begin{equation} \small
            \left[ \begin{NiceMatrix}
            J_{P_{ac} / E'} & J_{P_{ac} / E''}  & J_{P_{ac} / E_{dc}}   \\ 
            J_{Q_{ac} / E'} & J_{Q_{ac} / E''}  & J_{Q_{ac} / E_{dc}}    \\[2pt] 
            J_{E_{dc} / E'} & J_{E_{dc} / E''}  & J_{E_{dc} / E_{dc}}   \\
            J_{P_{ac}^+ / E'} & J_{P_{ac}^+ / E''}  & J_{P_{ac}^+ / E_{dc}}   \\
            J_{Q_{ac}^+ / E'} & J_{Q_{ac}^+ / E''}  & J_{Q_{ac}^+ / E_{dc}} \\[3pt]  
            J_{E_{ac}^{ 0\prime} / E'} & J_{E_{ac}^{0 \prime} / E''}  & J_{E_{ac}^{0 \prime} / E_{dc}}   \\
            J_{E_{ac}^{- \prime} / E'} & J_{E_{ac}^{- \prime} / E''}  & J_{E_{ac}^{- \prime} / E_{dc}}   \\
            J_{E_{ac}^{0 \prime \prime}/ E'} & J_{E_{ac}^{0 \prime \prime}/ E''}  & J_{E_{ac}^{0 \prime \prime}/ E_{dc}}   \\ 
            J_{E_{ac}^{- \prime \prime}/ E'} & J_{E_{ac}^{- \prime \prime}/ E''}  & J_{E_{ac}^{- \prime \prime}/ E_{dc}}   \\[3pt] 
            J_{P_{dc} / E'} & J_{P_{dc} / E''}  & J_{P_{dc} / E_{dc}}   
            \CodeAfter 
                \tikz \draw [dotted] (1-|3) -- (11-|3) ;
                \tikz \draw [dashed] (3-|1) -- (3-|8) ;
                \tikz \draw [dashed] (6-|1) -- (6-|8) ;
                \tikz \draw [dashed] (10.3-|1) -- (10.3-|8) ;
            \end{NiceMatrix} \right]
        \cdot 
            \left[  \begin{NiceMatrix}
            \Delta  E'  \\
            \Delta  E'' \\
            \Delta  E_{dc}   
            \CodeAfter 
                \tikz \draw [dotted] (3-|1) -- (3-|2) ;
            \end{NiceMatrix} \right]
            =
            \left[ \begin{NiceMatrix}
            \Delta P_{ac} \\
            \Delta Q_{ac} \\[2pt]
            \Delta E_{dc} \\
            \Delta P_{ac}^+ \\
            \Delta Q_{ac}^+ \\[3pt]
            \Delta E_{ac}^{0'} \\
            \Delta E_{ac}^{- \prime} \\
            \Delta E_{ac}^{0 \prime \prime} \\
            \Delta E_{ac}^{- \prime \prime} \\[3pt]
            \Delta P_{dc}
            \CodeAfter 
                \tikz \draw [dashed] (3-|1) -- (3-|8) ;
                \tikz \draw [dashed] (6-|1) -- (6-|8) ;
                \tikz \draw [dashed] (10-|1) -- (10-|8) ;
            \end{NiceMatrix} \right]\label{NR}
    \end{equation}
where e.g. $J_{P_{ac} / E'}$ is the partial derivative $\frac{\partial P_{ac}}{\partial E'}$.

The unknowns are updated at each step by taking the inverse of the Jacobian until the convergence criterium is reached.  
\begin{eqnarray}
    \mathbf{x}^{(\nu +1)} = \mathbf{x}^{(\nu)} + {\mathbf{J}^{(\nu)}}^{-1} \Delta \mathbf{y}^{(\nu)}
\end{eqnarray}
The convergence criterium is set on the update of the mismatches \eqref{conv} where $\epsilon$ is the desired tolerance:
\begin{eqnarray}
    \Delta \mathbf{y}^{(\nu)} < \epsilon \label{conv}
\end{eqnarray}

\section{Case study}
\label{Sec:Casestudy}

\begin{figure}
 \centering
   \includegraphics[width=0.8\linewidth]{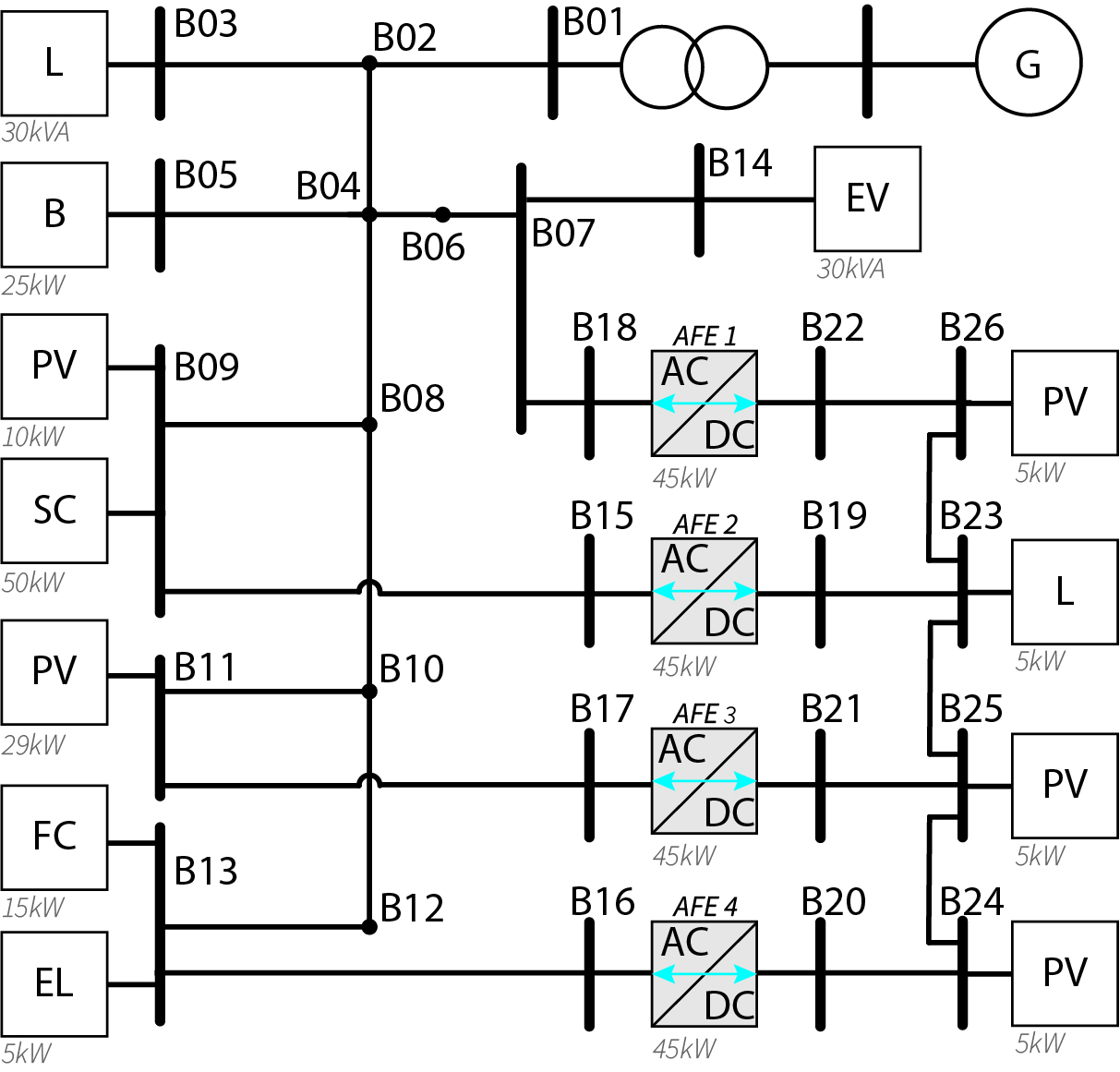}
   \caption{Topology of the hybrid AC/DC micro-grid developed at the EPFL}
   \label{fig:Mgrid}
 \end{figure}

The proposed PF algorithm for hybrid AC/DC is first validated on the hybrid AC/DC grid developed at the EPFL. The topology and parameters of the hybrid network are presented in  \cite{willem} and shown in Fig. \ref{fig:Mgrid}. The hybrid AC/DC microgrid consists of 18 AC nodes, 8 DC nodes, and 4 interfacing converters that can work under two control modes: $\mathbf{E_{dc} - Q_{ac}}$ and $\mathbf{P_{ac} - Q_{ac}}$. Tab. \ref{bustypes} summarises the node types in the hybrid network. Both grids have a base power of \SI{100}{\kilo VA} and a base voltage of \SI{400}{V_{ac}} and \SI{800}{V_{dc}}. Notice that any method previously proposed in the literature cannot be used for the PF analysis of this hybrid microgrid because two IC are controlling the DC voltage. The model is made publically available to the interested reader to reproduce the results at \url{https://github.com/DESL-EPFL}.

\renewcommand{\arraystretch}{1.2}
\begin{table}[!h] 
\centering
\caption{Node types in the hybrid AC/DC microgrid}
\begin{tabularx}{0.9\linewidth}{@{\extracolsep{\fill}} llllll }
 \multicolumn{2}{c}{\textbf{AC}} &  & \multicolumn{2}{c}{\textbf{DC}}   \\ \cline{1-2} \cline{1-2} \cline{4-5} \cline{4-5}
\textbf{Bus Type}      & \textbf{Bus \#} &  & \textbf{Bus Type}      & \textbf{Bus \#}   \\ \cline{1-2} \cline{4-5}
PQ            & 2-14   &  & P             & 24-26    \\ \cline{1-2} \cline{4-5}
VSC $(E_{dc} - Q_{ac} )$ & 15,18  &  & VSC $(E_{dc} - Q_{ac}) $ & 19,20    \\ \cline{1-2} \cline{4-5}
VSC $(P_{ac}$ - $Q_{ac})$   & 16,17  &  & VSC $(P_{ac}$ - $Q_{ac})$   & 21,22    \\ \cline{1-2} \cline{4-5}
AC slack         & 1      &  &               &         \\ \cline{1-2}
\end{tabularx} \label{bustypes}
\end{table}

\begin{figure}
\vspace{-0.5cm}
\centering
  \includegraphics[width=0.99\linewidth]{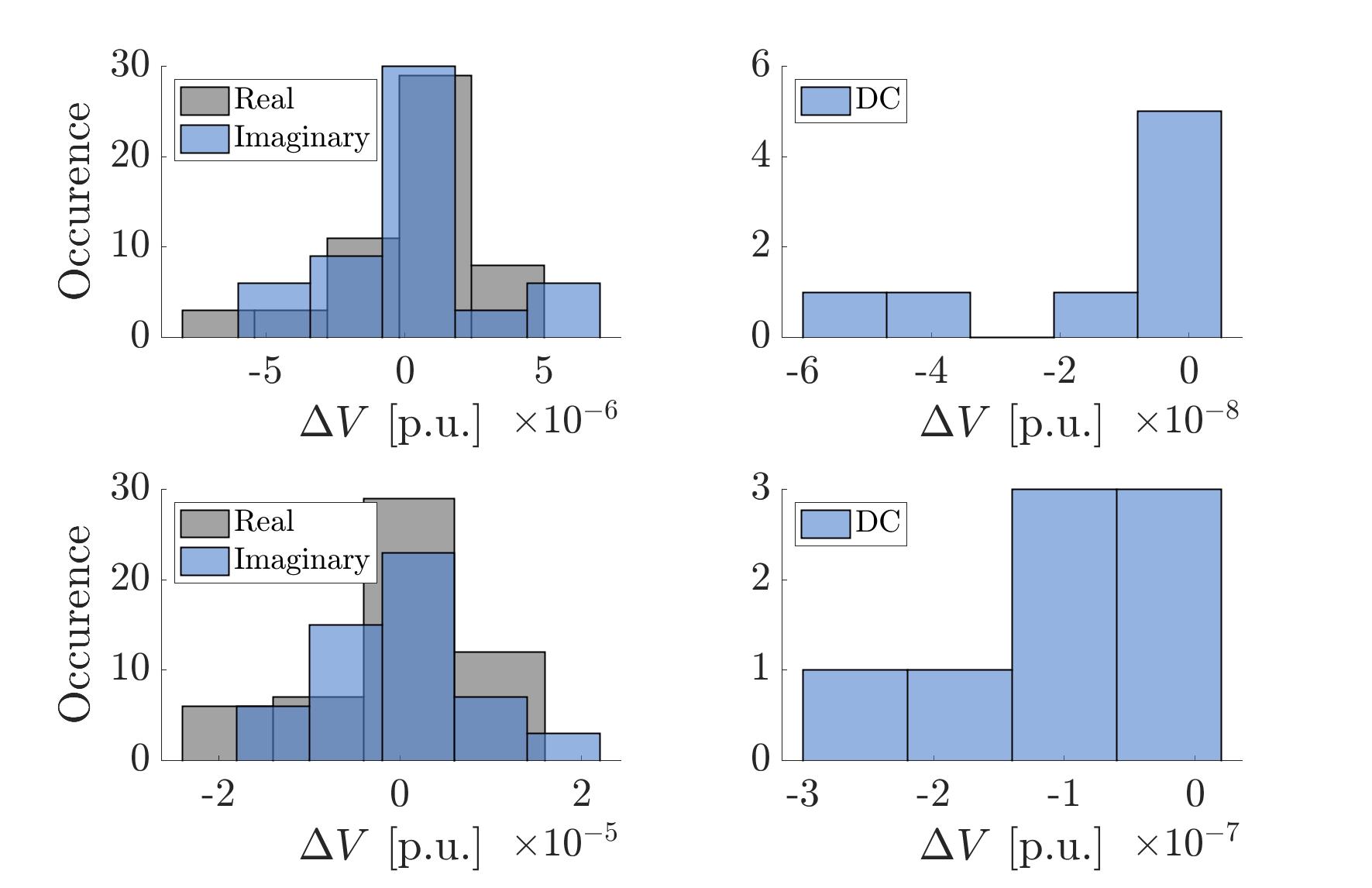}
  \caption{Histogram of the voltage errors between the true state and load flow results. Top - balanced grid, bottom - unbalanced grid. Left - AC grid, right - DC grid.}
  \label{Plot}
  \vspace{-0.4cm}
\end{figure}

Two steady-state time simulations are performed in the EMTP-RV simulation environment: a balanced one and a strongly unbalanced one, where the difference power injections between the phases in \textit{B09} reaches \num{0.5} p.u. The results from the simulation are considered the 'ground truth' and are used to validate the PF algorithm. The NR convergence criterium \eqref{conv} is set at $\epsilon < 10^{-6}$ for the update of the mismatches $\Delta \mathbf{y}$. The voltage errors, the difference between the ground truth and the results of the load flow algorithm, are presented as a histogram in Fig.\ref{Plot}. The top figures show the voltage errors of the real and imaginary components and the DC voltages for the balanced case. The results of the unbalanced case are shown in the two lower figures. For the balanced loading conditions, the maximum voltage error is in the order of \num{5e-6}. For the unbalanced loading conditions, the maximum error is in the order of \num{2e-5}. The NR algorithm takes around \SI{20}{\milli \second} and 4 iterations to converge for the considered three-phase 26-node hybrid microgrid when started from a \textit{flat start} where the voltage magnitudes are initialised at 1 p.u.


\section{Comparison with existing methods}
\label{sec:comparison}
The performance of the proposed method is also benchmarked against the FUBM-based PF algorithm proposed in \cite{alvarez2021flexible, alvarez2021universal}. The PF model has been made publicly available as an extension of the MATPOWER package \cite{zimmerman2010matpower}. As discussed in Section \ref{sec:stateoftheart}, the model is based on the universal branch model to model the ICs and requires four additional state variables to model each AC-DC interface: two variables ($m_a$and $\theta_a$) to model the amplitude modulation and the phase-shifting action of the PWM control of the VSC, a shunt susceptance $B_{eq}$ to compensate the reactive power injected into the DC grid and the variable $G^{sw}$, accounting for the VSC losses. The PF model is solved using the Newton-Raphson method.

The comparison is performed on four different hybrid AC/DC networks: 1) the hybrid microgrid developed at the EPFL presented in Section \ref{Sec:Casestudy}, 2) the IEEE 30-node system that has been extended with a 3 and 5-node DC network, 3) the IEEE 57-node network that is connected to the IEEE 14-node network using a DC network, and 4) the modified 1354-node PEGASE network to showcase the scalability of our proposed method.
The modifications made on the IEEE networks and the PEGASE network to include a DC system, have been proposed by \cite{alvarez2021flexible}.

The FUBM-based PF algorithm only works for balanced, single-phase systems (i.e. only the direct sequence equivalent is considered) and, furthermore, the method is limited to only one IC to control the DC voltage. Therefore, our proposed model has been appropriately adapted to consider only the direct sequence, and the studied networks only contain one voltage-controllable IC.

The two PF methods are compared in accuracy and computation time. 
To allow a fair comparison, the boundary conditions are set the same: the power setpoints of all the generators and loads, the type of nodes (\textit{PV} or \textit{PQ}) and the operating modes of the ICs. The convergence criteria \eqref{conv} for both methods are set at $\epsilon < 10^{-8}$ for the update of the mismatches. Furthermore, the initial conditions of the unknowns, $\mathbf{x}^{(0)}$, are all set at $1 p.u.$.

The results of the time analysis for the four reference hybrid AC/DC networks are presented in Table \ref{table:time_comp}. The CPU time (in seconds), the number of state variables, and the number of iterations required to reach the convergence criteria are shown. To allow a fair comparison, only the time to run the iterative NR process is considered, i.e., the computation of the mismatch equations, the construction of the Jacobian and the update of the unknowns. The columns \textit{EPFL} indicate the results of our proposed PF methods, while \textit{FUBM} refers to the results of the FUBM method presented in \cite{alvarez2021flexible}.

It can be seen that the computation time of the proposed method is one order of magnitude smaller than the FUBM method. This can be explained by the fact that the FUBM requires additional variables to model the ICs. Therefore, the number of unknowns increases along with the number of iterations and consequently, the computation time. The method proposed in this paper does not require additional variables to model the IC's behaviour since it is an extension of the AC-PF and only requires 2 variables for each AC node and 1 variable for each DC node. The large CPU time of the FUBM method is mainly due to the computation of the partial derivatives of the power injections with respect to the variables $m_a$ and $B_{eq}$ which are needed for the IC's model. 

The scalability of the method is demonstrated on the 1354-node PEGASE grid. Despite the large number of unknowns, the proposed method can solve the PF in 6 iterations in the order of seconds. The number of state variables is this time lower in the FUBM model. This is because the FUBM model is formulated in polar coordinates, and a large number of generators (261) are present in the PEGASE grid. The generators are modelled as PV nodes and because of the polar representation of the FUBM method, only one state variable per \textit{PV node} is needed. Our proposed method is formulated in rectangular coordinates and requires 2 state variables per PV node.


\begin{table}
\caption{Comparison of the computational time of the different PF methods.}
\label{table:time_comp}
\adjustbox{width=\columnwidth}{
\renewcommand{\arraystretch}{1.5}
\begin{tabular}{llllllllll}
           &  & \multicolumn{2}{c}{ \textbf{CPU time [s]}} &  & \multicolumn{2}{c}{\textbf{\# iterations}} &  & \multicolumn{2}{c}{\textbf{\# states}}\\ \cline{3-4} \cline{6-7} \cline{9-10}  
           &  & EPFL         & FUBM          &  & EPFL         & FUBM         &  & EPFL          & FUBM          \\ \hline
\textbf{$ \mu$ Grid (Sec. \ref{Sec:Casestudy})} &  & 0.018       & 0.515       &  & 4            & 11           &  & 41            & 60            \\
\textbf{IEEE 30}    &  & 0.023       & 0.412        &  & 4            & 5            &  & 82            & 109           \\
\textbf{IEEE 57+14}    &  & 0.035       & 0.469        &  & 4            & 5            &  & 174           & 200           \\
\textbf{PEGASE}     &  & 2.533       & 13.75       &  & 6            & 7            &  & 2724          & 2490          \\ \hline 
\end{tabular} }
\end{table}

\section{Conclusion}

In this paper, we present a novel model for the power flow in multiterminal hybrid AC/DC networks. The model is formulated in a general and unified way and solved using the Newton-Rapson method. New to previously published works, the proposed methodology allows multiple AC/DC converters to control the DC voltage. This is a crucial element in the planning and control of multiterminal AC/DC networks, as it corresponds more to the realistic operational condition of these hybrid grids. 
Additionally, the model is able to compute the power flow under balanced and unbalanced loading conditions and can account for intentionally negative sequence power injection. 

The method is numerically validated on a hybrid AC/DC microgrid, whose topology has been inspired by a real benchmark grid. Multiple ICs regulate the DC voltage level and the network is subjected to unbalanced loading conditions. It is shown that the error between the ground truth, obtained in an EMTP-RV time-domain simulation, and the results of the proposed PF method is very small in the order of \num{2e-5} to \num{5e-6}. 

Furthermore, the computational time of the proposed method is compared with the FUBM-based PF method, which is implemented as an extension of the MATPOWER tool. The comparison has been performed using multiple hybrid networks with different voltage levels, topologies, and sizes. The convergence criterion is set the same for both methods. It is shown that the method proposed in this paper converges by a factor of 10 faster than the FUBM-based method.


\bibliographystyle{IEEEtran}
\bibliography{IEEEexample.bib}

\end{document}